\documentclass[reprint]{revtex4-2}
\usepackage{graphicx}
\usepackage{dcolumn}
\usepackage{bm}
\usepackage[colorlinks=true, allcolors=blue]{hyperref}
\usepackage{siunitx}	
\usepackage{amsmath}
\usepackage{longtable}
\usepackage{booktabs}
\usepackage{multirow}
\usepackage{comment}
\usepackage{physics}

\usepackage[commandnameprefix=ifneeded]{changes}
\definechangesauthor[name={Natalia}, color=magenta]{no}
\definechangesauthor[name={Ben}, color=brown]{bo}
\definechangesauthor[name={Shikha}, color=blue]{SR}
\definechangesauthor[name={Igor}, color=orange]{iv}

\begin{document}

\preprint{APS/123-QED}

\title{Theory Framework for Medium-Mass Muonic Atoms}

\author{S.~Rathi}
\email[]{Corresponding author: shikha.rathi@campus.technion.ac.il}
\affiliation{The Helen Diller Quantum Center, Department of Physics, Technion--Israel Institute of Technology, Haifa, Israel\relax}

\author{I.~A.~Valuev}
\email[]{Corresponding author: igor.valuev@mpi-hd.mpg.de}
\affiliation{Max Planck Institute for Nuclear Physics, Saupfercheckweg 1, 69117 Heidelberg, Germany}

\author{Z.~Sun}
\affiliation{Max Planck Institute for Nuclear Physics, Saupfercheckweg 1, 69117 Heidelberg, Germany}

\author{M.~Heines}
\affiliation{KU Leuven, Instituut voor Kern- en Stralingsfysica, Leuven, Belgium}

\author{P.~Indelicato}
\affiliation{Laboratoire Kastler Brossel, Sorbonne Université, CNRS, ENS-PSL Research University, Collège de France, Paris, France}

\author{B.~Ohayon}
\affiliation{The Helen Diller Quantum Center, Department of Physics, Technion-Israel Institute of Technology, Haifa, Israel}

\author{N.~S.~Oreshkina}
\affiliation{Max Planck Institute for Nuclear Physics, Saupfercheckweg 1, 69117 Heidelberg, Germany\relax}

\begin{abstract}

We present a state-of-the-art theoretical approach for computing bound-state energies in muonic atoms, incorporating improved quantum electrodynamics effects and nuclear polarization corrections with a systematic assessment of theoretical uncertainties. 
Our approach is based on a combination of the $Z\alpha$-expansion and the all-order formalism (Furry picture) optimized for the medium-mass range $(3 \leq Z \lesssim 30)$ and guided by the accuracy requirements of modern muonic spectroscopy experiments.
These calculations are directly relevant to ongoing and forthcoming measurements aimed at extracting nuclear structure parameters, particularly nuclear charge radii, with unprecedented precision.
\end{abstract}

\maketitle


\section{\label{sec:Intro}Introduction}

Muonic x-ray spectroscopy is a powerful tool for probing nuclear structure~\cite{Borie, fricke} and for studying physics beyond the standard model~\cite{Nacy_paul_prl, okumura2023proof, beyer2025selfconsistentboundsstandardmodel, BEYER2024138746, lin2026probing, 2018-x17, 2026-NP}.
Many experiments have been performed in the past for a wide variety of reasons, from the determination of magnetic dipole, electric quadrupole, and higher multipole moments~\cite{povel1973m1, weber1982spectroscopic, powers1978measurement}, to testing QED~\cite{dubler1978precision} and studying muon cascade dynamics~\cite{schneuwly1993formation, kirch1999muonic}. However, the exploration of muon–nucleus bound-state dynamics in the past was limited by experimental constraints (detector efficiency and resolution) or by incomplete theoretical descriptions. The shortcomings on the theoretical side are emphasized in~\cite{ohayon2024critical,2025_Pb_prl, beyer2025relativisticrecoilkeyfinestructure}, where the reanalysis of previous muonic data with improved theory and careful uncertainty estimation yields a drastic change in nuclear charge radii value itself and/or its uncertainty.

Recent advances in measurement techniques, including cryogenic quantum sensors~\cite{ohayon2024towards, unger2024mmc} and spectroscopy with microgram-scale targets~\cite{adamczak2023muonic, antwis2025comparative}, are opening high-precision studies of muonic atoms to a broad range of elements, including both stable and radioactive species~\cite{ohayon2024towards, 2021-wauters-mux, 2025-rathi-reference, okumura2023proof}. 
%
\begin{figure}
    \centering
    \includegraphics[width=0.99\columnwidth]{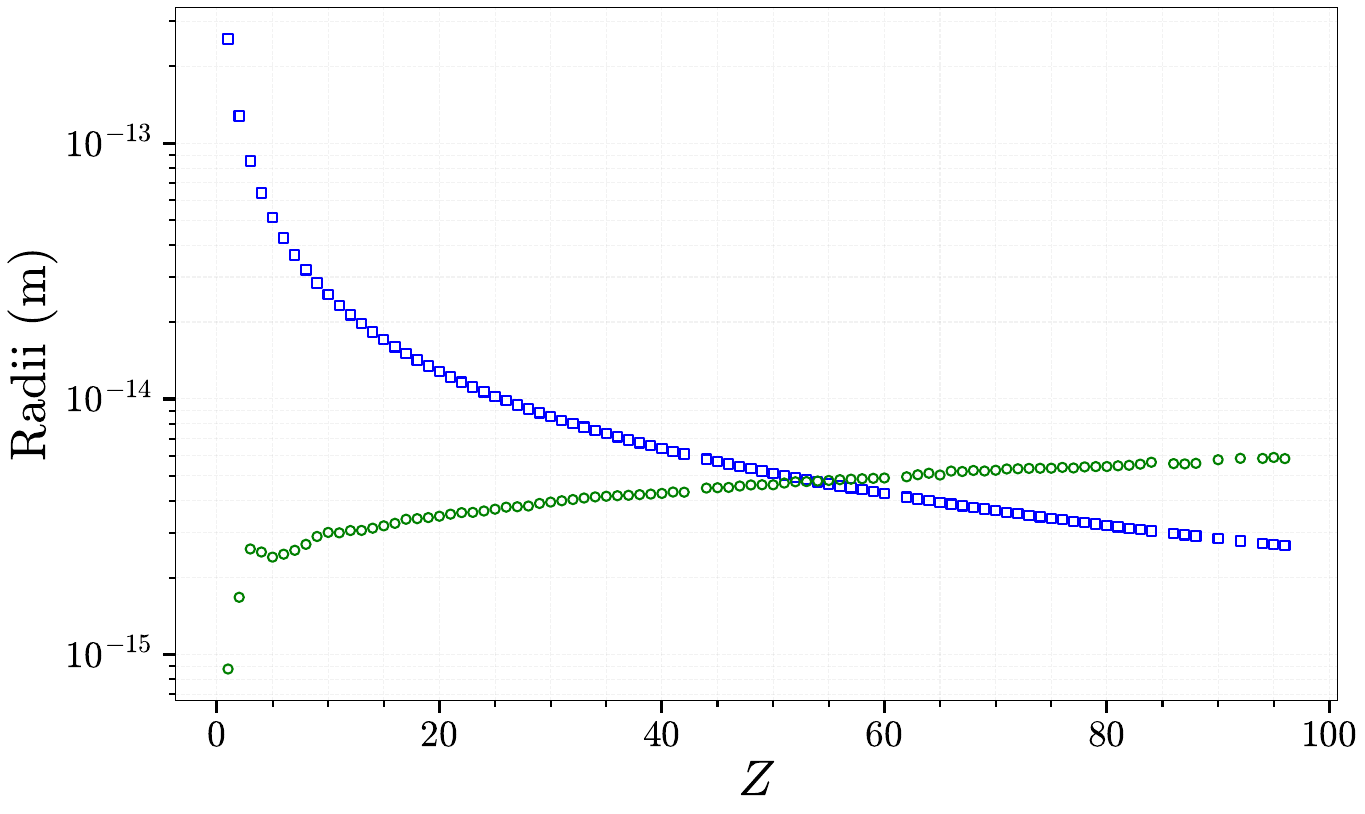}
    \caption{Comparison of Bohr-model of the \( 1s \) orbital radius (squares) and nuclear charge radii (circles) from~\cite{angeli2013table} for nuclei across the periodic table. 
    }
    \label{fig:radii_vs_Z}
\end{figure}
%

A crucial component for the success of these initiatives is the advancement of theoretical calculations. While the non-relativistic $Z\alpha$ perturbative approach has been successfully applied to muonic hydrogen ($Z=1$) and helium ($Z=2$) isotopes~\cite{2024-pachucki_comprehensive}, it is suboptimal for nuclei that are heavier than these due to the slow convergence of the perturbation series. For light atomic systems, the starting point is the Schrödinger equation for a point-like nucleus. Relativistic quantum electrodynamics (QED) and nuclear structure effects are incorporated within the non-relativistic quantum electrodynamics (NRQED) framework as a perturbative expansion in small parameters, such as the ratio of the nuclear charge radius to the Bohr radius (see Fig.~\ref{fig:radii_vs_Z}).

\begin{figure}
    \centering
    \includegraphics[width=0.99\linewidth]{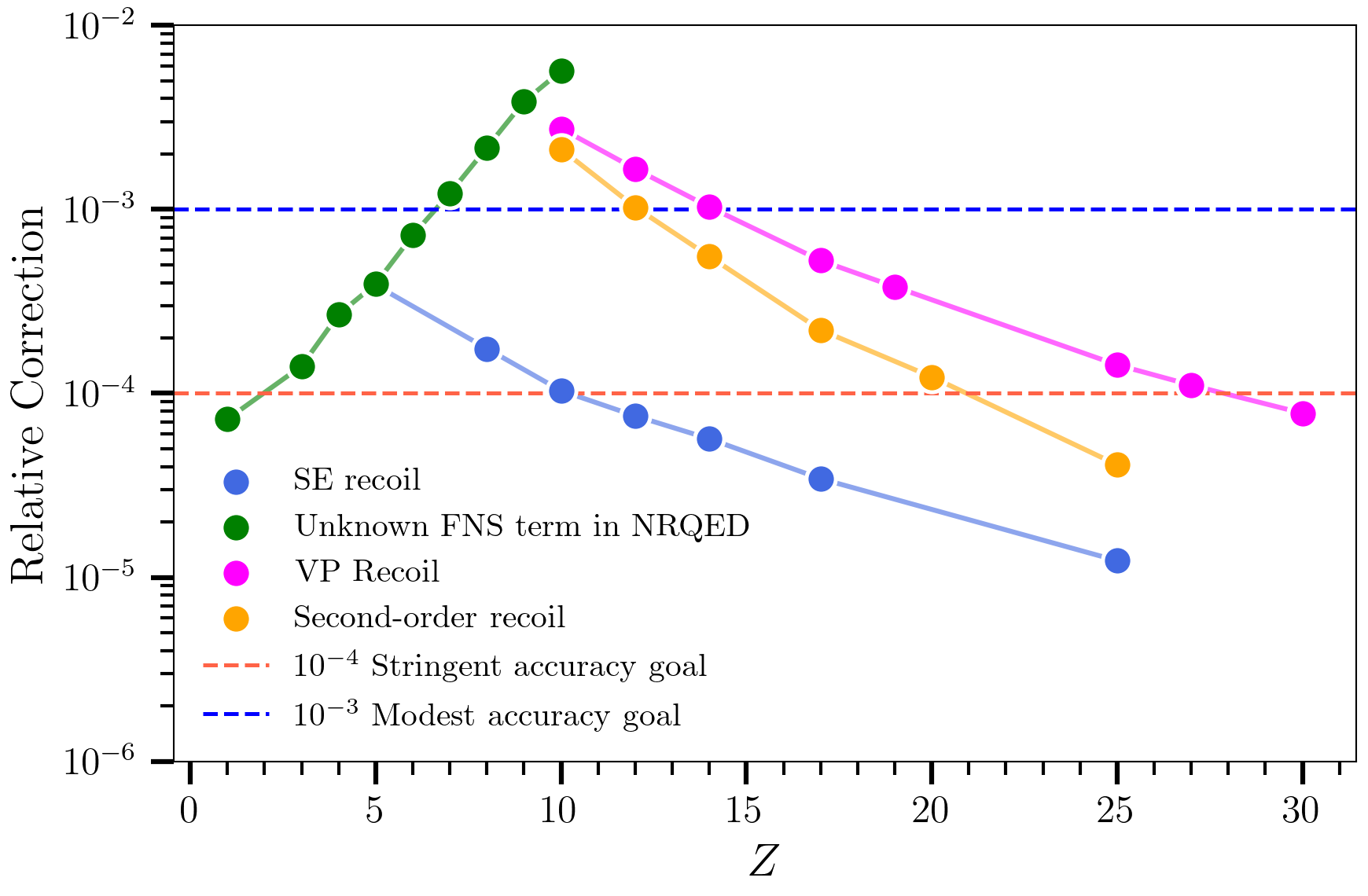}
    \caption{Corrections currently unknown in the NRQED and all-order approaches that are relevant to the $1s$ binding energies of muonic atoms, shown relative to the finite nuclear size contribution and compared with current experimental accuracy goals.}
    \label{missing and new correction} 
\end{figure}

An intuitive assessment of the validity of the perturbative $Z\alpha$-expansion approach to the finite-nuclear-size (FNS) correction is illustrated in Fig.~\ref{missing and new correction}. Here, we evaluate the FNS contribution using the known $Z\alpha$ expansion up to $(Z\alpha)^6$, following Ref.~\cite{2018_Pachucki}, for infinite nuclear mass. These perturbative results are compared with the all-order calculations obtained by solving the Dirac equation for the same nuclear charge distribution and subtracting the corresponding point-nucleus solution.

The difference between the Dirac-based result and the truncated $(Z\alpha)^6$ expansion isolates the missing higher-order contributions in NRQED, starting at $\mathcal{O}\!\bigl((Z\alpha)^7\bigr)$.
In view of ongoing experimental efforts~\cite{ohayon2024towards, beyer2025modern} targeting a relative accuracy of $10^{-3}-10^{-4}\times \delta E_\text{FNS}$, where $\delta  E_\text{FNS}$ is the finite size contribution, the unknown $Z\alpha$ expansion
terms become significant around $Z=3-7$, depending on the accuracy goal.

On the other hand, if we move towards the higher $Z$, the ratio of the nuclear charge radius to the muonic $1s$ Bohr radius increases rapidly. A qualitative illustration is shown in Fig.~\ref{fig:radii_vs_Z}.
Evidently, for $Z \gtrsim 40$, the muon is localized very close to or even within the nuclear volume.
In this regime, the Dirac equation must be solved in a fully self-consistent framework that incorporates the nuclear and electronic fields. All-order theoretical treatments for such heavy systems have advanced significantly in recent years~\cite{oreshkina2022self, yerokhin2023qed, mandrykina2025hadronic, mandrykina2026wichmann, sommerfeldt2026all}. 

In the intermediate region, $3 \le Z \lesssim 30$, both perturbative and all-order approaches are, in principle, applicable, each with its own advantages and limitations. 
For instance, the current all-order methods suitable for heavy elements are limited to linear nuclear recoil corrections~\cite{yerokhin2023qed}. 
However, for intermediate-$Z$ nuclei, higher-order recoil contributions become non-negligible (see Fig.~\ref{missing and new correction}) and may limit the theoretical accuracy when compared to current experimental precision goals.
Conversely, in this regime, the muon-to-nuclear mass ratio ($m_\mu/m_{\rm nucl}$) remains sufficiently small to justify a systematic expansion-based treatment of higher-order recoil effects.

Despite the overlapping advantages offered by these two theoretical formalisms, no single framework has emerged as the standard in this regime, highlighting the need for a unified treatment that combines their respective strengths.
Therefore, in this work, we construct such a hybrid theoretical framework by merging the $Z\alpha$ expansion with all-order methods. As a first application, we present our calculations for muonic \(^{35,37}\mathrm{Cl}\) ($Z=17$), for which new measurements have recently been performed. Together with the experimental data, these calculations enable improved determinations of both absolute and differential nuclear charge radii. Details of the experimental methodology and extracted more precise radii are presented in a companion publication~\cite{beyer2025modern}.

This article is arranged as follows: In section~\ref{sec: all-order}, we begin by discussing the known all-order approach to evaluate the muonic atom binding energies. 
We discuss the nuclear polarization correction (NP) and its uncertainty estimation in section~\ref{sec:NP}. 
In section~\ref{sec: advancement} we introduce new corrections based on the all-order and $\alpha Z$ expansion approaches. 
Unless otherwise stated, atomic units are used throughout this work.

\section{\label{sec: all-order} All order approach}

In general, the transition energies ($E_{\mathrm{exp}}$) measured in muonic spectroscopy experiments can be theoretically described by decomposing the total energy into a series of well-defined contributions as follows,
\begin{equation}
E_{\mathrm{exp}} =
E_{\mathrm{point}} +
  \delta E_{\mathrm{FNS}}
+ \delta E_{\mathrm{QED}}
+ \delta E_{\mathrm{recoil}}
+ \delta E_{\mathrm{nucl}} \, .
\end{equation}
Here, $E_{\mathrm{point}}$ denotes the energy calculated within the point-nucleus approximation. 
$\delta E_{\mathrm{QED}}$, $\delta E_{\mathrm{recoil}}$, and $\delta E_{\mathrm{nucl}}$ represent the corrections arising from the QED effects, finite nuclear mass (recoil), and additional nuclear-structure effects such as nuclear polarization and deformation, respectively. 

For muonic atoms, the dominant correction to the binding energy originates from the finite nuclear size. In fact, in medium- and high-$Z$, the FNS effect is sufficiently large that it is generally not separated from the point-nucleus contribution. Instead, the point-nucleus value itself is taken as the reference (zero) contribution.
Consequently, our framework incorporates the finite nuclear size at the zeroth order level by solving the Dirac equation in the electrostatic potential $V_{\rm nucl}(r)$ of an extended nuclear charge distribution in the infinite nuclear mass limit.
%
\begin{equation}\label{eq:theo_Dirac}
    \left[ \boldsymbol{\alpha} \cdot \mathbf{p} + \beta m_{\mu} + V_{\text{nucl}}(r) \right] | \psi_{n\kappa m} \rangle = E_{n\kappa} | \psi_{n\kappa m} \rangle,
\end{equation}
where $| \psi_{n\kappa m} \rangle$ and $E_{n\kappa}$ are the eigenfunction and eigenvalues, respectively, for a bound state with quantum numbers $n$ (principal), $\kappa$ (relativistic angular momentum), and $m$ (magnetic). $\boldsymbol{\alpha}$ and $\beta$ are the Dirac matrices, $\mathbf{p}$ is the muon momentum operator, and $m_{\mu}$ is the muon mass.
    
In the particular case of chlorine, we used a two-parameter Fermi distribution to model the nuclear charge distribution, which can be given as:
\begin{equation}\label{eq:theo_Fermi2}
        \rho(r) = \frac{\rho_0}{1 + \exp\!\left(\frac{r - c}{a}\right)}.
    \end{equation}
Here, $\rho_0$ is the normalization factor, and $c$ is a nuclear size parameter.
The surface diffuseness parameter $a$ is related to the skin thickness $t$ through
\begin{equation}
t = 4 \ln(3)\, a.
\end{equation}

In the present region of interest, we fix $t = 2.3~\si{\femto\meter}$, which is a common assumption for this region~\cite{fricke}. 
For heavier systems, where the sensitivity to the nuclear surface becomes more pronounced, 
both $c$ and $a$ should be treated as free parameters in a proper fit~\cite{beyer2025relativisticrecoilkeyfinestructure, 2025_Pb_prl}.

\subsection{QED corrections}
The next dominant corrections originate from QED effects, namely vacuum polarization (VP) and self-energy (SE). In muonic atoms, vacuum polarization constitutes the leading QED contribution owing to the small orbital radius of the muon, which strongly enhances short-range vacuum effects, whereas self-energy contributions remain comparatively small. 

\begin{figure}
    \centering
    \includegraphics[width=\linewidth]{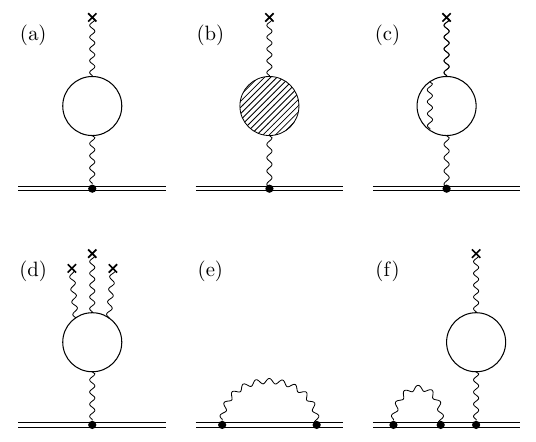}
    \caption{Feynman diagram representation of the QED contributions considered in this work. (a) One-loop Uehling VP (VP$_{11}$), (b) Hadronic VP (hVP$_{11}$), (c)   K\"all\'en-Sabry  VP (eVP$_{11}^2$), (d) Wichmann-Kroll VP (VP$_{13}$),  (e) Self-Energy (SE),  (f) Interlinked SE-VP (SE-eVP). For illustrative purposes, only a single representative diagram is shown for each contribution class.}
    \label{fig:loop_dia_for_each_correction}
\end{figure}

The leading-order VP term, illustrated in Fig.~\ref{fig:loop_dia_for_each_correction}(a), is described by the Uehling potential and is of order $\mathrm{eVP}_{11} \propto \alpha (Z\alpha)$~\cite{uehling1935polarization,klarsfeld1977analytical}. One may account for $\text{eVP}_{11}$ perturbatively, with Dirac-Coulomb wavefunctions, or incorporate it directly into the potential in Eq.~\eqref{eq:theo_Dirac}. In our framework, we choose the latter option as it provides higher-order, non-negligible contributions from reducible two-loop corrections (loop-after-loop effects) and also yields Dirac-Coulomb-Uehling (DCU) wavefunctions that can be used to better approximate the perturbative contributions calculated next.
Hence, the potential in Eq.~\eqref{eq:theo_Dirac} can be replaced by a total potential as follows: 
\begin{equation}\label{eq:theory_potential}
    V(r) = V_\text{nucl}(r) + e\text{VP}_{11}(r).
\end{equation}
Substituting Eq.~\eqref{eq:theory_potential} into Eq.~\eqref{eq:theo_Dirac}, the Dirac equation is solved self-consistently. 
In this work, we used the multiconfiguration Dirac Fock and general matrix elements (MDFGME) code~\cite{indelicato2024mcdfgme}, while alternative approaches such as the dual-kinetic-balance (DKB) method~\cite{dkb} with a piecewise-polynomial representation are also suitable and yield similar results. 

Beyond the zeroth-order approximation discussed above, further VP-related phenomena that affect the binding energies included in this work are as follows:
(1) The muonic vacuum polarization contribution (Fig.~\ref{fig:loop_dia_for_each_correction}(a)), arising from virtual $\mu^+\mu^-$ pairs, evaluated using the Uehling potential with the muon mass. 
(2) The hadronic vacuum polarization correction (Fig.~\ref{fig:loop_dia_for_each_correction}(b)), incorporated following the methodology of Ref.~\cite{breidenbach2022hadronic}, extended here to muonic atoms with a Fermi nuclear charge distribution. 
(3) The irreducible two-loop electron vacuum polarization  contribution (Fig.~\ref{fig:loop_dia_for_each_correction}(c)), described by the K\"all\'en--Sabry potential~\cite{huang1976calculation}. 
(4) The next-to-leading-order correction in $Z\alpha$ to the one-loop eVP (Fig.~\ref{fig:loop_dia_for_each_correction}(d)), commonly referred to as the Wichmann--Kroll (WK) contribution~\cite{wichmann1956vacuum,huang1976calculation}, evaluated following the procedure detailed in Ref.~\cite{sommerfeldt2026all}.

All these effects can be treated either perturbatively or by incorporating them into the effective potential (Eq.~\eqref{eq:theory_potential}) and subsequently solving the Dirac equation~Eq.~\eqref{eq:theo_Dirac} with this modified potential. Since their contributions are small compared with the leading-order term, both approaches yield equivalent results at the level of accuracy relevant here. We therefore consider only corrections larger than $1~$ppm 
of the leading-order energy. This accuracy goal is more than sufficient to ensure that the extracted radii will be limited only by nuclear structure effects, not by missing QED corrections.

Additional QED contributions include the self-energy (see Fig.~\ref{fig:loop_dia_for_each_correction} (e)), which cannot be represented by an effective potential. In our framework, we calculate the SE, including the finite-size correction, following Ref.~\cite{oreshkina2022self}. The associated uncertainty in the SE-FNS contribution is estimated by accounting for variations across nuclear-charge-distribution models and numerical convergence. Similar calculations of SE with finite size corrections are performed by P. Mohr et al.~\cite{SE_FNS_Paul} 
as an extension to their method discussed in Ref.~\cite{mohr1993nuclear} for muonic atoms. The results are consistent within the quoted uncertainties.

\subsection{Relativistic recoil}\label{rel_recoil}
Until now, we have treated the nucleus as infinitely heavy in the bound-state calculation; corrections arising from its finite mass must be included. These nuclear recoil effects are evaluated within a rigorous QED framework to first order in the mass ratio $m_\mu / m_{\rm nucl}$, following Ref.~\cite{2023-uRec} for a two-parameter Fermi nuclear charge distribution. This procedure yields a fully relativistic recoil correction evaluated to all orders in $Z\alpha$.

In Table~\ref{tab:energy_cl35_37}, we present our calculations for each of the above discussed contributions for $1s$ and $np$, ($n=2\text{–}4$) levels of $\mu^{35}$Cl. The corresponding values for the recoil corrections for $\mu^{37}$Cl can be obtained by using the mass ratios.

\subsection{Electron screening}

For completeness and to directly complement our sister publication on the experimental details, we evaluate the electron screening effect by considering fully occupied $1s$ and $2s$ electronic orbitals. The total screening at the individual level is substantial (of the order $\si{\kilo\eV}$). However, most of it cancels when considering transitions.
The calculated effects are $-$0.41\,eV, $-$1.12\,eV, and $-$1.53\,eV for the $2p\to1s$, $3p\to1s$, and $4p\to1s$ transitions, respectively.

\begin{table*}[ht]
\centering
\caption{Total binding energy and contributions from different corrections for $\mu$\,$^{35}$Cl $1s_{1/2}$ level. The results are for $r_\mathrm{nucl} =3.380~\si{\femto\meter}$  with $t =2.3~\si{\femto\meter}$. 
}
\begin{ruledtabular}
\begin{tabular}{l l r r r r r r r
}
\multicolumn{1}{c}{} & \multicolumn{1}{c}{} & \multicolumn{7}{c}{Energies (eV)} \\
\cline{3-9}
 & Contribution & $1s_{1/2}$\phantom{(0)}  & $2p_{1/2}$   & $2p_{3/2}$   & $3p_{1/2}$   & $3p_{3/2}$    & $4p_{1/2}$   & $4p_{3/2}$    \\
     \midrule
$\delta E_{\rm FNS}$ & $E(\mathrm{V_C+eVP_{11}})$               & $-783~978$\phantom{(0)} & $-204~796$ & $-204~015$& $-90~836$& $-90~606$& $-51~033$& $-50~936$\\
$\delta E_{\rm QED}$ &~ ~ $\Delta E(\mathrm{\mu VP_{11}})$     & $-8$\phantom{(0)}      & $0$       & $0$      & $0$      & $0$     & $0$     & $0$          \\
&~ ~ $\Delta E(\mathrm{hVP_{11}})$        & $-5$\phantom{(0)}      & $0$       & $0$      & $0$      & $0$     & $0$     & $0$          \\
&~ ~ $\Delta E(\mathrm{eVP^2_{11}})$      & $-38$\phantom{(0)}     & $-4$      & $-4$     & $-1$     & $-1$     & $-1$     & $-1$       \\
&~ ~ $\Delta E(\mathrm{eVP_{13}})$        & $3$\phantom{(0)}       & $1$        & $1$       & $0$     & $0$       & $0$     & $0$               \\
&~ ~ $\Delta E(\mathrm{SE})$              & $150(4)$  & $-1$      & $1$        & $0$     & $0$     & $0$     & $0$       \\
&~ ~ $\Delta E(\mathrm{SE-eVP}) $         & $2(2)$    & $-$      & $-$    & $-$   & $-$  & $-$  &  $-$    \\
$\delta E_{\rm recoil}$&~ ~ $\Delta E_{\mathrm{rec}}^{(1)}$      & $2~330$\phantom{(0)}   & $661$    & $659$       & $294$   & $293$   & $165$   & $165$      \\
&~ ~ $\Delta E_{\mathrm{VP_{rec}}}^{(1)}$ & $20$\phantom{(0)}     & $3$      & $3$    & $1$   & $1$  & $0$  &  $0$      \\      
&~ ~ $\Delta E(\mathrm{SE_{rec}}) $       & $-1$\phantom{(0)}      & $-$      & $-$    & $-$   & $-$  & $-$  &  $-$    \\
&~ ~ $\Delta E_{\mathrm{rec}}^{(2)}$      & $-7$\phantom{(0)}      & $-$      & $-$    & $-$   & $-$  & $-$  &  $-$    \\
\midrule
& Total                                & $-781~532(4)$ & $-204~136$ & $-203~355$ & $-90~542$ & $-90~313$ & $-50~869$ & $-50~772$\\
\end{tabular}
\end{ruledtabular}
\label{tab:energy_cl35_37}
\end{table*}

\section{Nuclear polarization}\label{sec:NP}
While the reduced distance to the nucleus is the prime benefit of muonic atoms, it also amplifies the effects of the internal dynamic nuclear structure beyond the finite size. 
These effects are commonly referred to as nuclear polarization. 
In this section, we present an approach to NP that combines well-established techniques with recent advancements in microscopic calculations in order to gain a better insight into theoretical uncertainties. This approach is intended for application to deformed nuclei in both medium-mass and heavy muonic atoms, and we demonstrate the calculations for $^{35,37}\text{Cl}$ as an example.

First of all, since in the medium- and high-$Z$ regime the FNS must be incorporated to all orders in $Z\alpha$ from the very beginning (see Section~\ref{sec: all-order}), one needs a framework that treats NP as a perturbation with respect to the non-perturbative FNS as a starting point.
This is achieved by representing the total nuclear four-current density operator $\hat{J}^{\mu}_{\text{N}}(x)$ as the following sum:
\begin{align}
\label{eq:J_nucl}
\hat{J}^{\mu}_{\text{N}}(x) = J^{\mu}_{\text{N},\,\text{stat}}(\boldsymbol{\mathrm{x}}) + \hat{J}^{\mu}_{\text{N},\,\text{fluc}}(x),
\end{align}
where the classical static part $J^{\mu}_{\text{N},\,\text{stat}}(\boldsymbol{\mathrm{x}})$ corresponds to the average over the nuclear ground state, and the remaining fluctuating part $ \hat{J}^{\mu}_{\text{N},\,\text{fluc}}(x)$ defines the perturbation due to the intrinsic nuclear dynamics. 
The static part is taken into account at the level of the Dirac equation in the form of the potential $V_{\rm nucl}(r)$ (see Section~\ref{sec: all-order}, Eq.~\eqref{eq:theo_Dirac}) and ultimately gets absorbed into the dressed muon-propagator, whereas the fluctuating part can be treated by associating with it an additional second-quantized photon field $\hat{A}^{\mu}_{\text{fluc}}(x)$.
This way, after combining this new term with the usual free-photon operator $\hat{A}^{\mu}_{\text{free}}(x)$ into the total radiation field
\begin{equation}
\hat{A}^{\mu}_{\text{rad}}(x) =\hat{A}^{\mu}_{\text{free}}(x)+\hat{A}^{\mu}_{\text{fluc}}(x),
\label{eq:Radiative field}
\end{equation}
one is led to a modified photon propagator as follows:
\begin{align}
\mathcal{D}_{\mu\nu}(x,x') & = -i \bra{0} T[ \hat{A}_{\mu}^{\text{rad}}(x)  \hat{A}_{\nu}^{\text{rad}}(x')] \ket{0} \notag \\
& = D_{\mu\nu}^{\text{free}}(x-x') + D^{\text{NP}}_{\mu\nu}(x,x'),
\label{eq:Modified propagator}
\end{align}
where $D_{\mu\nu}^{\text{free}}$ is the free photon propagator, and $D_{\mu\nu}^{\text{NP}}$ is the correction due to NP.

\begin{figure}[!tbp]
\centering
\includegraphics[scale=1]{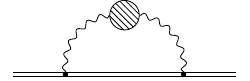}
\caption{Leading-order NP as effective self-energy. 
The shaded circle represents the NP insertion.}
\label{fig:SE-NP}
\end{figure}

In such a framework, the NP effect is seamlessly incorporated into QED, with the only addition that now every photon line receives a correction in the form of the so-called NP insertion.
A diagrammatic depiction of the corresponding leading-order effective self-energy diagram is shown in Fig.~\ref{fig:SE-NP}, and we refer to~\cite{1989_Plunien, 1991_Plunien, Haga_e_2002, 2024_Valuev} for more details on the formalism.
We also note that a similar one-loop diagram involving vacuum polarization does not need to be taken into account since it has been shown to be effectively absorbed in the measured value of the nuclear charge radius~\cite{2024_Valuev}.

NP represents the most challenging and uncertain correction to muonic energy levels, as the evaluation of the NP insertion involves summation over the entire nuclear spectrum of the isotope of interest. 
The summation over nuclear excitations can be divided into three contributions: 1)~Low-lying states (LLS), 2)~High-frequency collective excitations known as giant resonances (GR), and 3)~Excitations in the hadronic range corresponding to the so-called nucleon polarization (nP). 
The nP has only recently been considered for systems beyond \textsuperscript{4}He~\cite{gorchtein2026guide}, and it provides a shift equivalent to half of the NP uncertainty for the chlorine isotopes. The distinction between LLS and GR is made due to the different ways of taking them into account when fully microscopic calculations of a nuclear spectrum are not available, which is the case for most odd-mass nuclei, including \textsuperscript{35}Cl and \textsuperscript{37}Cl. The nuclear parameters for LLS are taken from nuclear data sheets~\cite{Nucl_data_35, Nucl_data_37}, while in the case of GR one has to resort to phenomenological energy-weighted sum rules~\cite{rinker1978nuclear}. In such cases, one also needs to choose an appropriate model for nuclear transition charge densities, which is of major importance for muonic atoms. 

In this work, nuclear transitions with multipolarities up to $L = 5$ are included in the calculations for muonic $^{35,37}$Cl.
For the Coulomb part of the muon-nucleus interaction, the nuclear transition charge densities are described by~\cite{2007_Haga}: 1) the monopole Tassie model for $L = 0$, 2) the hydrodynamical Jensen-Steinwedel model for $L = 1$, and 3) the Tassie-Goldhaber-Teller model for $L \geq 2$.
The models were chosen based on a comparison with microscopic Skyrme calculations~\cite{skyrme_rpa} for the nearby doubly-magic \textsuperscript{40}Ca, from which one can also expect that the corrections obtained for $^{35,37}\text{Cl}$ are most likely somewhat underestimated. 
The transverse contribution to NP was estimated by means of the continuity equation for nuclear charge and current transition densities, with the $(J, J+1)$ component of the current set to zero~\cite{2007_Haga}. The reliability of such estimations was controlled by performing the calculations in both Feynman and Coulomb gauges in order to check the fulfillment of gauge invariance. The resulting NP corrections for individual energy levels of muonic $^{35,37}\text{Cl}$ are listed in Table~\ref{tab:NP_shifts}. NP increases the binding energy (indicated by the negative sign), which leads to larger energies for $np\to1s$ transitions.

    \begin{table}
        \centering
        \caption{Nuclear part of NP (in $\si{\eV}$, Coulomb + transverse, up to $L=5$) for the energy levels of muonic $^{35,37}\text{Cl}$. Extra digits are shown in order to better demonstrate the differences between the levels and isotopes.}
        \begin{ruledtabular}
        \begin{tabular}{c|ccccccc}
            Isotope & $1s$\textsubscript{1/2} & $2p$\textsubscript{1/2} & $2p$\textsubscript{3/2} & $3p$\textsubscript{1/2} & $3p$\textsubscript{3/2} & $4p$\textsubscript{1/2} & $4p$\textsubscript{3/2} \\
            \hline
            \textsuperscript{35}Cl & $-$103.91 & $-$0.67 & $-$0.65 & $-$0.22 & $-$0.21 & $-$0.10 & $-$0.09 \\
            \textsuperscript{37}Cl & $-$100.05 & $-$0.63 & $-$0.61 & $-$0.20 & $-$0.20 & $-$0.09 & $-$0.09
        \end{tabular}
        \label{tab:NP_shifts}
        \end{ruledtabular}
    \end{table}

For the uncertainty evaluation, the adopted approach of using the simple non-microscopic charge transition densities was benchmarked against microscopic Skyrme calculations for \textsuperscript{40}Ca. 
Similar to Ref.~\cite{valuev2022evidence}, a representative set of Skyrme parameterizations was selected in such a way that they 1) Cover significant portions of the constraints on various nuclear saturation properties and 2) Represent different groups and fitting protocols. 
A comparison between the approximation applied in this work and different Skyrme models is shown in Table~\ref{tab:NP_skyrmes}.
Based on these results, the relative uncertainty was conservatively estimated by taking the maximally deviating Skyrme model (SkM*), giving an uncertainty of $(209.8 - 170.5)/170.5 \approx 23\%$.
For the nucleon part, the values were taken from Ref.~\cite{gorchtein2026guide} ($\sim$10\% uncertainty). When adding the different contributions, their uncertainties were added linearly as a conservative estimate. 
The resulting NP corrections to the muonic transition energies are given in Table~\ref{tab:NP}, along with the nucleon polarization taken from Ref.~\cite{gorchtein2026guide}. Given that NP is dominated by the $1s$ state, it is nearly identical for all $np\to1s$ transitions.
It should be noted that in this formalism, the next-order NP correction is negligible compared to the systematic uncertainty, as it corresponds to an additional loop and is suppressed by a factor of $\alpha$. 
Such subleading effects as the inelastic three-photon exchange considered in NRQED for light systems~\cite{2018_Pachucki} cannot be disentangled here due to the fact that our starting point already includes a full-order treatment in $Z\alpha$.

    \begin{table*}[]
        \centering
        \caption{Nuclear part of NP (in eV, Coulomb only, up to $L=3$) for the $1s_{1/2}$ state of muonic \textsuperscript{40}Ca calculated using different Skyrme models as well as the approach adopted for muonic $^{35,37}\text{Cl}_{17}$.}
        \begin{ruledtabular}
        \begin{tabular}{c|ccccccccccc}
            Approach adopted for $^{35,37}\text{Cl}$ & LNS & SKI3 & KDE0 & SKX & SLy5 & BSk14 & SAMi & NRAPR & SkP & SkM\textsuperscript{*} & SGII \\
            \hline
            $-$170.5 & $-$197.9 & $-$187.7 & $-$188.4 & $-$194.5 & $-$195.3 & $-$194.6 & $-$198.5 & $-$192.3 & $-$207.0 & $-$209.8 & $-$207.8
        \end{tabular}
        \end{ruledtabular}
        \label{tab:NP_skyrmes}
    \end{table*}

    \begin{table}
        \centering
        \caption{Calculations of NP (in $\si{\eV}$) for the relevant transitions in \textsuperscript{35}Cl and \textsuperscript{37}Cl. $f_{\text{Corr}}$ (\%) denotes the fraction of NP that behaves in a correlated way between the two isotopes.}
        \begin{ruledtabular}
        \begin{tabular}{cc|cc|cc}
            Isotope & Transition & Nuclear & Nucleon~\cite{gorchtein2026guide} & Total & $f_{\text{Corr}}$ (\%) \\
            \hline
            \textsuperscript{35}Cl \rule{0pt}{2.5ex} & $2p1s$ & 103(24) & 11.9(1.2) & 115(25) & 90.4 \\
                                   & $3p1s$ & 104(24) & 11.9(1.2) & 116(26) & 90.4 \\
                                   & $4p1s$ & 104(24) & 11.9(1.2) & 116(26) & 90.4 \\
            \hline
            \textsuperscript{37}Cl \rule{0pt}{2.5ex} & $2p1s$ & 99(23)  & 12.6(1.3) & 112(25) & 97.6 \\
                                   & $3p1s$ & 100(23) & 12.6(1.3) & 112(25) & 97.6 \\
                                   & $4p1s$ & 100(23) & 12.6(1.3) & 113(25) & 97.6 \\
        \end{tabular}
        \end{ruledtabular}
        \label{tab:NP}
    \end{table}

When studying muonic isotope shifts, part of the NP uncertainty cancels out due to correlations, primarily originating from the GR and nP contributions. In the treatment by Fricke and Heilig~\cite{fricke}, the uncertainty of the NP difference is assumed to be 10\% of the larger NP value. Rather than directly assigning an uncertainty to the NP difference, we approximate the correlation between the NP contributions of the two isotopes. The GR and nP terms are strongly correlated between isotopes, whereas the LLS contribution is largely uncorrelated. Accordingly, we assume maximal correlation for the GR and nP components and no correlation for the LLS component. Ideally, the difference in each of these components would then be considered separately, but a reliable uncertainty estimate for the individual nuclear components is highly non-trivial. Instead, we define the \textit{correlated fraction} $f_{\text{Corr}}$ of the NP as the part originating from the GR and nP contributions. The correlation between the NP contributions of the two isotopes is then estimated as the product of their correlated fractions, yielding 88.2\%. This corresponds to a NP difference of \textsuperscript{37}Cl relative to \textsuperscript{35}Cl of $-3(12)~\si{\eV}$. Notably, the uncertainties obtained with this approach are very similar (though not by construction) to those derived using the prescription of Fricke and Heilig~\cite{fricke}.
    
\section{\label{sec: advancement} New QED corrections specific to Medium-$Z$}

Beyond the established contributions discussed above, the present work introduces several new theoretical developments that improve the accuracy and consistency of the calculated energy levels. In this section, we outline these advancements and quantify their impact.

\subsection{\label{sec:Non Rel Recoil}Recoil Correction to VP }
As mentioned earlier, at the level of precision targeted by modern muonic spectroscopy experiments (see Fig.~\ref{missing and new correction}),  it is no longer sufficient to treat the nucleus as infinitely heavy. The finite nuclear mass induces recoil corrections to the QED contribution.
Although suppressed by the mass ratio $m_\mu/m_\mathrm{nucl}$, these recoil correction terms for VP can reach the eV level due to enhanced VP effects in muonic atoms. Therefore, calculating these corrections is relevant for high-precision determinations of nuclear-structure parameters. 

We evaluate the leading-order recoil correction to vacuum polarization (VP$_{\mathrm{rec}}$) following the formalism of Ref.~\cite{yerokhin2023qed}. Specifically, we modify the potential in Eq.~(8) of Ref.~\cite{yerokhin2023qed} by including the Uehling potential, such that the leading order recoil correction $E_L$ reads,
\begin{equation}\label{VPrec_equation}
E_{L} = \langle a|(\boldsymbol{\alpha}\!\cdot\!\mathbf{p})^{2}|a\rangle 
= \langle a|(E_a - \beta m_\mu - V(r))^2|a\rangle ,
\end{equation}
where $V(r)$ is defined in Eq.~\eqref{eq:theory_potential}. $E_a$ is the Dirac energy of eigenstate $a$.  The inclusion of the Uehling potential, together with the nuclear Coulomb potential, yields the leading linear recoil contribution evaluated to all orders in $Z\alpha$, including VP effects. The recoil correction to VP is then obtained as the difference between the solutions of Eq.~(\ref{VPrec_equation}) with and without the Uehling potential.

This approach does not constitute a fully rigorous QED treatment, as VP recoil contributions from two-photon exchange are not included. However, the leading VP$_{\mathrm{rec}}$ term is sufficient at the present level of experimental and theoretical accuracy. The relative magnitude of VP$_{\mathrm{rec}}$ with respect to the leading FNS and the relevant $Z$ regime is shown in Fig.~\ref{missing and new correction}. It becomes non-negligible for $Z<30$ with stringent accuracy goals and for $Z<16$ with modest accuracy goals.


\subsection{\label{sec: expansion method} 
Corrections calculated in the nonrelativistic approximation}

\subsubsection{Second-order recoil}



While the linear recoil contribution is included to all orders in $Z\alpha$ within the fully relativistic framework described in the previous section~\ref{rel_recoil}, we employ an expansion in the mass ratio $m_\mu/m_{\rm nucl}$ to isolate the quadratic recoil term proportional to $(m_\mu/m_{\rm nucl})^2$. In practice, the all-order recoil contribution is first evaluated, and the linear recoil term is subsequently subtracted, allowing the quadratic component to be identified. The robustness of this procedure is verified by benchmarking against calculations for a point nucleus and by comparing with the expected behavior in the small-radius expansion. For medium-mass systems, where $m_\mu/m_{\rm nucl} \ll 1$, the expansion remains rapidly convergent and provides a controlled method for separating higher-order recoil contributions.

To this end, we begin by specifying the recoil in the nonrelativistic limit for a point nucleus, where recoil effects are fully accounted for by replacing the muon mass with the reduced mass of the system. The binding energy of a muonic level for a finite nuclear mass can therefore be written as
\begin{align}\label{quad recoil}
E_{m_\mathrm{red}} = \frac{m_\mathrm{red}}{m_\mu}\, E_{m_\mu},
\end{align}
where $m_\mathrm{red} = \frac{m_\mu m_{\mathrm{nucl}}}{m_\mu + m_{\mathrm{nucl}}}$ denotes the reduced mass. Expanding Eq.~\eqref{quad recoil} in powers of the mass ratio $m_\mu/m_{\rm nucl}$ yields
\begin{align}\label{eq:NR energy recoil}
\frac{E_{m_\mathrm{red}}}{E_{m_\mu}}
&= a_0 
+ a_1 \left(\frac{m_\mu}{m_{\rm nucl}}\right)
+ a_2 \left(\frac{m_\mu}{m_{\rm nucl}}\right)^2  \nonumber \\
&\quad + \mathcal{O}\!\left[\left(\frac{m_\mu}{m_{\rm nucl}}\right)^3\right],
\end{align}
where the coefficients $a_i$ define the recoil contributions of successive orders in the mass-ratio expansion.
For the non-relativistic reduced-mass relation of Eq.~\eqref{quad recoil}, the coefficients are simply $a_0 = 1$, $a_1 = -1$, and $a_2 = 1$. The terms proportional to $(m_\mu/m_{\rm nucl})$ and $(m_\mu/m_{\rm nucl})^2$ correspond to the linear and quadratic recoil corrections, respectively.

To numerically extract the recoil coefficients, the ratio $E_{m_\mathrm{red}}/E_{m_\mu}$ is evaluated for $\mu\mathrm{Cl}_{17}$ over a range of nuclear masses, corresponding to different values of $m_\mu/m_{\rm nucl}$. The resulting data is fitted with a high-order polynomial (typically up to 12th order) in $m_\mu/m_{\rm nucl}$, allowing for a precise determination of the coefficients $a_i$. The agreement between the analytically expected values ($a_0 = 1$, $a_1 = -1$, $a_2 = 1$) and the numerically extracted coefficients confirms the numerical stability of our recoil extraction procedure (see Table~\ref{tab:recoil_coefficients}).

As a benchmark for extending this analysis to a nucleus of finite size, we first adopt the small-radius approximation, which allows for a direct comparison with an analytical result:
\begin{align}\label{eq:finite_recoil_small_r}
E_{r, m_\mathrm{red}} =
\left(\frac{m_\mathrm{red}}{m_\mu}\right) E_{m_\mu}
+ \left(\frac{m_\mathrm{red}}{m_\mu}\right)^3
F\,r_{\rm nucl}^2 ,
\end{align}
where $F\,r_{\rm nucl}^2$ denotes the leading FNS contribution. For the $1s$ state in the infinite nuclear-mass limit, the coefficient $F$ is given by $F = \frac{2}{3} Z\alpha^4 m_\mu$. Substituting the Taylor expansion of the reduced mass in Eq.~\ref{eq:finite_recoil_small_r} yields
\begin{align}
\label{eq:FNS_recoil_expansion}
\frac{E_{(r,m_\mathrm{red})}}{E_{m_\mu}}
&= \left[1 + \frac{F\,r_{\rm nucl}^{2}}{E_{m_\mu}}\right] \nonumber \\
&\quad - \left(\frac{m_{\mu}}{m_{\rm nucl}}\right)
\left[1 + 3 \frac{F\,r_{\rm nucl}^{2}}{E_{m_\mu}}\right] \nonumber \\
&\quad + \left(\frac{m_{\mu}}{m_{\rm nucl}}\right)^{2}
\left[1 + 6 \frac{F\,r_{\rm nucl}^{2}}{E_{m_\mu}}\right]
+ \mathcal{O}\!\left[\left(\frac{m_\mu}{m_{\rm nucl}}\right)^3\right].
\end{align}
We validate our numerical extraction method by comparing the fitted coefficients with those obtained from Eq.~\eqref{eq:FNS_recoil_expansion} (see Table~\ref{tab:recoil_coefficients}). The excellent agreement confirms the robustness of the fitting procedure.

\begin{table*}[htb]
\caption{Comparison of analytical and numerical recoil coefficients for the
non-relativistic expansion in $m_\mathrm{red}/m_{\mu}$ for $\mu\text{Cl}_{17}$.}
\centering
\begin{tabular}{l|c|c|c|c|c|c|c|c|c}
\hline\hline
\multirow{2}{*}{\textbf{Case}} 
    & \multicolumn{3}{c|}{$0^{\text{th}}$ Order ($a_0$)}
    & \multicolumn{3}{c|}{$1^{\text{st}}$ Order ($a_1$)}
    & \multicolumn{3}{c}{$2^{\text{nd}}$ Order ($a_2$)} \\
\cline{2-10}
& Analytical & Numerical & \% diff   
& Analytical & Numerical & \% diff
& Analytical & Numerical & \% diff \\
\hline
Point Nucleus ($r_\mathrm{nucl}=0$) 
& 1 &1.000000 & 
& $-$1 & $-1.000000$ & 
& 1 & 1.000000 & \\
Small $r_\mathrm{nucl}$ limit ($r_\mathrm{nucl}=0.1$ fm)
&    0.999941 &    0.999942 & 6.83E$-05$
& $-$0.999824 & $-$0.999826 & 2.73E$-04$
&    0.999647 &    0.999654 & 6.67E$-04$ \\
$r_\mathrm{nucl}$ ($r_\mathrm{nucl}=3.380$ fm)
&    
&    0.955686 & 
& 
& $-$0.885215 & 
&             &    0.801966 & - \\
\hline\hline
\end{tabular}
\label{tab:recoil_coefficients}
\end{table*}

Applying the above procedure to a nuclear radius $r_{\rm nucl}=3.380~\mathrm{fm}$, close to the physical charge radius of Cl, we again determine the recoil coefficients from a polynomial fit in $m_\mu/m_{\rm nucl}$. In this case, the zeroth-order coefficient corresponds to the Dirac binding energy for an infinite nuclear mass, including the finite nuclear-size correction. 
The quadratic recoil contribution is obtained from the second-order coefficient $a_2$ of the fit.

Following this, for $\mu^{35}\mathrm{Cl}_{17}$, the extracted coefficient $a_2 = 0.802$ yields the second-order recoil contribution as,
\begin{equation}
\Delta E_{\text{recoil}}^{(2)} 
= E_{m_\mu}\left(\frac{m_{\mu}}{m_{\rm nucl}}\right)^2 a_{2}
= -6.9~\mathrm{eV}.
\end{equation}

To highlight the need for the quadratic recoil contribution in the interpretation of muonic atom spectroscopy experiments, we compare its magnitude for selected nuclear charges with the targeted experimental accuracy (see Fig.~\ref{missing and new correction}). The quadratic recoil term exceeds the stringent projected precision goal for atoms with $Z \lesssim 20$, demonstrating that second-order recoil corrections must be included in theoretical predictions for muonic atom energies in this regime.

 \subsubsection{Recoil Correction to SE}
 In a non-relativistic scenario with a point-nucleus, the  recoil correction to the leading-order one-loop self-energy can be expressed for the SE$_{\text{rec}}$, for the $1s$ state, in first-order in $m_\mu/m_{\rm nucl}$ as 
 \begin{align}
 \Delta E_\text{SE} & \approx 
 - 2 \frac{\alpha (Z\alpha)^4 m_\mu}{\pi} 
 \Bigl(\frac{m_\mu}{m_{\rm nucl}}\Bigr) \times \nonumber \\ 
 & \Biggl[ 
 -4\ln(Z\alpha) + 2\ln{k_0(1,0)} 
 + 1
 \Biggr].
 \end{align}

 %
 Here, ln\;$k_0(1,0)$ is the Bethe logarithm for $n=1$ and $l=0$. 
 As seen in Fig.~\ref{missing and new correction}, it is negligible compared with the accuracy goals for all but the lightest elements, where the approximations made to it are justified.
 In Cl, it returns a negligible $-1.4\,$eV, which is expected to have a finite-size correction of order $30\%$.

\subsubsection{SE-eVP}
This is the combination of one loop of SE and one of eVP (see Fig.~\ref{fig:loop_dia_for_each_correction} (f)).
For the $1s$ state in a point-like muonic atom, the natural magnitude of this correction $(\alpha/\pi)^2(Z\alpha)^4\,m_\mu$ , which tends to $1\,$ppm for large $Z$. However, at least in light systems, the numerical coefficient multiplying it is of the order of $20-30$\,\cite{2024-Se-eVP}.
Substituting $Z=17$ for Cl results in a rough upper magnitude of $\Delta E(\mathrm{SE-eVP})\approx3.5\,$eV. This number is expected to obtain a finite-size correction that reduces it by about $30\%$, so we adopt a conservative estimate of $\Delta E(\mathrm{SE-eVP})\approx2(2)\,$eV.


\section{\label{conclusion}Conclusion}
A new generation of experimental campaigns aimed at high-precision spectroscopy of muonic atoms across the periodic table is currently underway or being proposed, placing increasingly stringent demands on theoretical input.
Motivated by this need, we have developed a unified theoretical approach for medium-mass systems ($3 \leq Z \lesssim 30$) that combines the complementary strengths of the $Z\alpha$ expansion and all-order methods while overcoming their respective limitations. This synthesis enables the inclusion of previously neglected effects, such as linear recoil correction to vacuum polarization and self-energy, and non-relativistic second-order recoil. 
With this advance, we establish a new benchmark for muonic atom theory and open new avenues for experimental programs probing bound-state QED and nuclear structure at unprecedented precision.



\section*{Acknowledgments}
S. R. gratefully acknowledges the Technion and the Davis Postdoctoral Fellowship. S.R. carried out part of this work at the Max Planck Institute for Nuclear Physics, Heidelberg, supported by the Max Planck–Israel Postdoctoral Visiting Programme. N.S.O. thanks the DFG (German Research Foundation) – Project-ID 273811115 – SFB 1225 ISOQUANT for funding. 

\bibliography{bib}

\end{document}